\documentclass{article}

\usepackage{graphicx}
\usepackage{amsmath}
\usepackage{longtable}
\usepackage{booktabs}
\usepackage{multirow}
\usepackage{float}

\usepackage[dblblindworkshop, final]{neurips_2026}

\usepackage[utf8]{inputenc} 
\usepackage[T1]{fontenc}    
\usepackage{hyperref}       
\usepackage{url}            
\usepackage{booktabs}       
\usepackage{amsfonts}       
\usepackage{nicefrac}       
\usepackage{microtype}      
\usepackage{xcolor}         

\usepackage{booktabs}
\usepackage{tabularx}
\usepackage{hyperref}

\title{VitisGP: Multi-Task Benchmark for Genotype to Phenotype Grapevine Prediction}

\workshoptitle{Tackling Climate Change with Machine Learning}
\author{%
  Chiara D'Ercoli \\
  Department of Computer, Control and Management Engineering \\
  Sapienza University of Rome\\
  \texttt{cd956@cam.ac.uk} \\
  \And
  May Levin \\
  Department of Computer Science \\
  University of Cambridge \\
  \texttt{ml2229@cam.ac.uk} \\
  \And
  Pietro Liò \\
  Department of Computer Science \\
  University of Cambridge \\
  \texttt{pl219@cam.ac.uk} \\
}

\begin{document}

\maketitle

\begin{abstract}
As climate change reshapes grape-growing conditions, predicting grapevine phenotypes from genomic variation is increasingly important for identifying and breeding cultivars suited to emerging environmental pressures. Existing research on the grapevine genotype–phenotype relationship using machine learning has often focused on individual traits, isolated regions, or a limited set of predictive models.
We introduce VitisGP, a multi-phenotype curated dataset, integrating genome-wide SNP data with 26 heterogeneous phenotype prediction tasks spanning sommelier-curated aroma traits (N=11), botanical traits (N=14) and wine color---across 1,197 Vitis vinifera cultivars. We employ VitisGP to benchmark phenotype prediction, comparing machine-learning models with differing inductive biases---L2-regularized logistic regression, Random Forest, XGBoost, CatBoost, GBLUP and Transformer-based models, providing the first systematic comparison of this breadth on harmonized grapevine data. 
Across the benchmark, CatBoost achieves the strongest overall performance (macro-average AUROC=$0.696$ across 26 tasks), although model performance varies substantially across phenotype classes, with sensory and color traits showing stronger predictability than agronomic characteristics. VitisGP provides the community with a resource for investigating the genomic basis of phenotypic traits and screening grapevine diversity for climate-relevant characteristics.
\end{abstract}

\section{Introduction}
The accelerating pace of climate change poses a growing threat to global agriculture, with grapevine (Vitis vinifera) cultivation being particularly vulnerable. 
In recent years, unpredictable temperature regimes, water availability, and disease pressure have severely affected grapevine growth and, in extreme cases, compromised entire cultivars \cite{wolkovich2018pinot,climateaffectswine}. 
Such environmental pressures directly alter grape physiology with downstream consequences for grape and wine quality \cite{vanleeuwen2019adaptation, climateaffectswine}. These challenges, together with the growing need for climate-resilient cultivars, motivate the development of computational approaches linking grapevine genetic variation to agronomic and sensory phenotypes.

Machine learning (ML) has increasingly been applied to agriculture as genomic and chemical measurements from various crops---grapes, wheat, and other produce---become available at larger scales.  Recent deep-learning approaches in genomics \cite{enformer, dnabert}, as well as methods developed specifically for crop genomics such as CropFormer \cite{cropformer} and GEFormer \cite{geformer}, have explored convolutional and attention-based architectures to learn representations of genomic variation. In grapevine, a central objective is to characterize the genotype-phenotype relationship, mainly identifying genetic traits/variation associated not only with specific sensory characteristics, but also with traits relevant to resilience under changing environmental conditions.

With more than 10,000 named cultivars, V. vinifera encompasses a wide array of agronomic, sensory, and terroir-adaptive traits. Current work, however, has largely focused on relatively narrow phenotype definitions or on isolated regions: the study proposed in \cite{predictingwineunderchangingclimate} predicts wine quality from climate data in Slovenian regions \cite{bayeswine} examines only grapevine population structure; \cite{genophenotablegrapes} investigates genomic-botanical associations in wine and table grapes; \cite{mlgeneralwinequality} trains classifiers on alcohol and acidity as quality proxies and \cite{mlpinotnoirquality} focuses on a single cultivar. These also include predicting aroma precursors \cite{flutre2022prediction}, and drought-related physiological traits \cite{brault2021drought}.  While these studies demonstrate the feasibility of computational phenotype prediction, they remain limited in task diversity, cultivar set size and model capacity.

More widespread wine review datasets \cite{spawn99}, provide valuable free-text descriptions, but tend to pool user data without a common protocol. Often, academic studies also employ inconsistent standards for chemical and sensory characterization \cite{yangWineBehavior}.  At the same time, professional wine-tasting frameworks---such as those developed by the Wine \& Spirit Education Trust (WSET) \cite{wset} --- provide standardized vocabularies for describing aroma, structure, and other sensory characteristics. Combining information available in existing wine resources with the structured terminology of professional tasting frameworks, offers an opportunity to construct richer representations of grapevine phenotypes. 
We therefore assemble a dataset of 1,197 Vitis vinifera cultivars with sommelier-profiled olfactory descriptors, structural attributes and botany-level agronomic traits, paired with genome-wide Single Nucleotide Polymorphism (SNP) data. To our knowledge, our dataset is the first to combine high-fidelity, expert-curated sensory phenotypes with grapevine genomics.
Using VitisGP, we propose a benchmark to investigate the following question: Across heterogeneous grapevine phenotypes, which class of predictive model---linear, tree-based ensemble, or attention-based---provides the most reliable genotype-to-phenotype prediction, and does the answer vary by phenotype class?

Our contributions are threefold: first, we present a curated grapevine dataset of SNP genotype data from 1,197 V. vinifera cultivars paired with high-fidelity sommelier sensory profiles and agronomic phenotypes. Second, we compare a diverse set of established prediction methods, including L2-regularized logistic regression (L2LR), GBLUP, elastic net, LASSO, Random Forest (RF), XGBoost (XGB), CatBoost (CB), and Transformers. Third, we explore the stability and recurrence of selected SNPs across folds and phenotype tasks, highlighting the most predictive markers and genomic regions.

\section{Methods}
\textbf{Dataset} Our study integrates publicly available genomic data with curated phenotypes to construct a multi-phenotype benchmark for genomic prediction. VitisGP is published and available at https://zenodo.org/records/22081041.

\textbf{Genotype data} Single Nucleotide Polymorphism (SNP) data were derived from the grapevine diversity panels reported by Laucou et al.~\cite{lacou}, D'Onofrio et al.~\cite{donofrio}, and De Lorenzis et al.~\cite{delorenzis}. Each panel comprises a collection of cultivated grapevine varieties and was originally characterized using genome-wide molecular markers. For our study, we organized the data by cultivar in a genotype matrix, comprising $1,197$ cultivars$\times 10{,}206$ SNP markers (783 from Laucou et al. \cite{lacou}, 312 from D'Onofrio et al. \cite{donofrio} and 102 from De Lorenzis et al. \cite{delorenzis}). Each representative was characterized using the same ordered catalog of 10,206 SNPs mapped to the PN40024 12X.V0 reference genome, with genotypes encoded as 0, 1, or 2 and missing calls retained explicitly.

\textbf{Phenotypes}
Phenotypic information was assembled from complementary sources to capture grapevine variation, including aroma and sensory characteristics, and botanical and agronomic traits including morphological and phenological cultivar properties \cite{spawn99} (for additional details refer to Table \ref{tab:web-data-sources} in the Appendix).
The sensory component was constructed from expert-curated descriptions of grape and wine characteristic, including the WSET lexicon \cite{wset}. Additionally, we derived a categorical phenotype describing grape color, represented as a binary prediction target---available for the complete panel of 1,197 cultivars. To complement the sensory phenotype space, we included botanical and agronomic characteristics, distinct from sensory perception. The original heterogeneous botanical annotations were harmonized into standardized traits by mapping to a structured common representation. The final dataset includes 26 traits, all reported in Table \ref{tab:benchmark-tasks}. Details on the preprocessing steps are in the Appendix \ref{phenotypes}.

\begin{table}[h]
\caption{Composition of the grapevine genotype-to-phenotype benchmark.
$N$ denotes the number of cultivar identities with an available label for
each prediction task.}
\label{tab:benchmark-tasks}
\centering
\footnotesize
\setlength{\tabcolsep}{2.5pt}
\renewcommand{\arraystretch}{1.0}
\begin{tabular}{
    p{1.3cm}
    p{5.5cm}
    p{0.6cm}
    p{1cm}
    p{4.5cm}}
\toprule
\textbf{Block} &
\textbf{Prediction targets} &
\textbf{Tasks} &
\textbf{$N$} &
\textbf{Example labels} \\
\midrule

Color &
Red color family &
1 &
1,197 &
\{non-red, red\}. \\

Aroma &
Floral, citrus fruit, red fruit, green fruit, pungent spice,
black fruit, stone fruit, herbaceous/herbal, mineral,
dried/cooked fruit, and tropical fruit &
11 &
398--472 &
Each aroma family: \{absent, present\}. \\

Botanical/ agronomic &
Titratable acidity, berry size, berry weight, budburst timing,
ripening period, maturity timing, bunch size, bunch compactness,
powdery mildew resistance, yield, fertility, Botrytis resistance,
must sugar level, and downy mildew resistance &
14 &
250--753 &
Berry size: \{small, medium, large\};
budburst timing: \{early, medium, late\};
titratable acidity: \{low, medium, high\};
bunch compactness: \{loose, medium, compact\}, [...]. \\
\bottomrule
\end{tabular}
\end{table}

\textbf{Models}
We evaluate models spanning three algorithmic families: tree-based models (RF, XGB, CB), an L2-regularized logistic regression---which all operate on the same marker feature set (10,206 SNPs)---and two Transformer configurations \cite{attnnnn}. The latter are both framed as a genomic predictor that represents each SNP marker as an independent categorical token, enriched with learned marker-identity and genomic-location embeddings. The two configurations encompass a single-task Transformer---trained independently for each phenotype, and a multi-task Transformer---that instead shares the encoder across all tasks to learn correlated phenotypic representations. We also include conventional GBLUP fitted by REML, elastic net and
LASSO, using the same fold assignments (Appendix~\ref{genomic-baselines}).

\textbf{Genotype Preprocessing \& Marker Selection.}
Each cultivar is initially represented by the complete set of SNP markers. 
In order to reduce marker noise and dimensionality, for the Transformer models we utilize a genomic-prediction strategy of univariate SNP pre-filtering, in which within each training fold, markers are ranked by their marginal association with the phenotype in case, and a fixed number ($k\!=\!256$) of the highest-ranked ones is retained~\citep{roshan2011ranking}. 
Tokenization information, architectural details and ablations results (choice of the number of k-features among ${128, 512, 2048, [...]}$), are in \ref{transformer}. 

\textbf{Loss, Training and Evaluation.}The tree-based models and L2 logistic regression use class balancing,
while the Transformer models are optimized with class-weighted cross-entropy loss. Per-task class weights are computed as the ratio of total training samples to per-class sample counts, then normalized by their mean and label smoothing ($\varepsilon\!=\!0.01$). All models except the Transformers use fixed-capacity protocols (details in \ref{training}). 
We employ a five-fold identity-group-disjoint cross-validation protocol. All cultivar identities are partitioned into three non-overlapping sets per fold (train/test/val), and they never appear in more than one partition within a fold, preventing information leakage (\ref{evaluation}).

\section{Results}
As reported in Table \ref{tab:macro_model_performance}, CatBoost achieves the strongest overall performance while Random Forest is the most consistently strong model across individual tasks. Both single-task Transformer and multi-task Transformer trail behind the tree-based models despite their architectural complexity. These results suggest that in moderate-sample genomic prediction regimes, simpler models with strong feature-interaction capacity can outperform deep architectures designed to model long-range genomic dependencies.
Botanical traits yield more conservative performance than the color and aroma tasks across all models. The best-performing model on berry weight achieves approximately 0.748 (RF), whereas results for yield level 0.618 (RF) and bunch compactness 0.657 (RF) are substantially lower. This suggests that highly epistatic traits (e.g., overall yield) may require larger sample sizes to resolve, whereas traits with stronger main-effects are more consistently predictable.
For the aroma prediction tasks---which have smaller task-specific sample sizes and significant class imbalance---the best models demonstrate strong capacity to map SNP genotypes to complex sensory vocabularies. Citrus fruit 0.921 (CB), red fruit 0.890 (CB), and black fruit 0.871 (CB) are the most predictable aroma families across model classes. 

\begin{table}[h]
\centering
\renewcommand{\arraystretch}{1.0}
\caption{Macro-averaged model performance across phenotypic trait categories. Values are AUROC/BAC; $N$ counts tasks with the highest mean AUROC among the nine models.}
\label{tab:macro_model_performance}
\footnotesize
\begin{tabular}{lccccc}
\toprule
Model & Color & Aroma & Botany & Overall & $N$ \\
\midrule
CB
& \textbf{0.908} / 0.878
& \textbf{0.753} / 0.642
& 0.637 / 0.446
& \textbf{0.696} / \textbf{0.545}
& 5 \\
XGB
& 0.898 / 0.873
& 0.746 / 0.636
& 0.632 / 0.440
& 0.691 / 0.539
& 0 \\
RF
& 0.895 / 0.854
& 0.716 / 0.563
& 0.650 / 0.406
& 0.687 / 0.490
& 12 \\
MT-Transformer
& 0.889 / 0.863
& 0.744 / \textbf{0.667}
& 0.591 / 0.412
& 0.667 / 0.537
& 2 \\
ST-Transformer
& 0.895 / 0.876
& 0.694 / 0.634
& 0.592 / 0.405
& 0.647 / 0.520
& 0 \\
L2-LR
& 0.820 / 0.747
& 0.666 / 0.609
& 0.624 / \textbf{0.447}
& 0.649 / 0.527
& 0 \\
Elastic net
& 0.895 / \textbf{0.880}
& 0.741 / 0.652
& 0.619 / 0.394
& 0.681 / 0.522
& 0 \\
GBLUP
& 0.825 / 0.759
& 0.669 / 0.582
& \textbf{0.652} / 0.409
& 0.666 / 0.496
& 5 \\
LASSO
& 0.895 / 0.877
& 0.737 / 0.647
& 0.613 / 0.398
& 0.677 / 0.522
& 2 \\
\bottomrule
\end{tabular}
\end{table}

\subsection{Post-hoc SNP analysis}

\begin{figure}[h]
    \centering
    \includegraphics[width=0.9\linewidth]{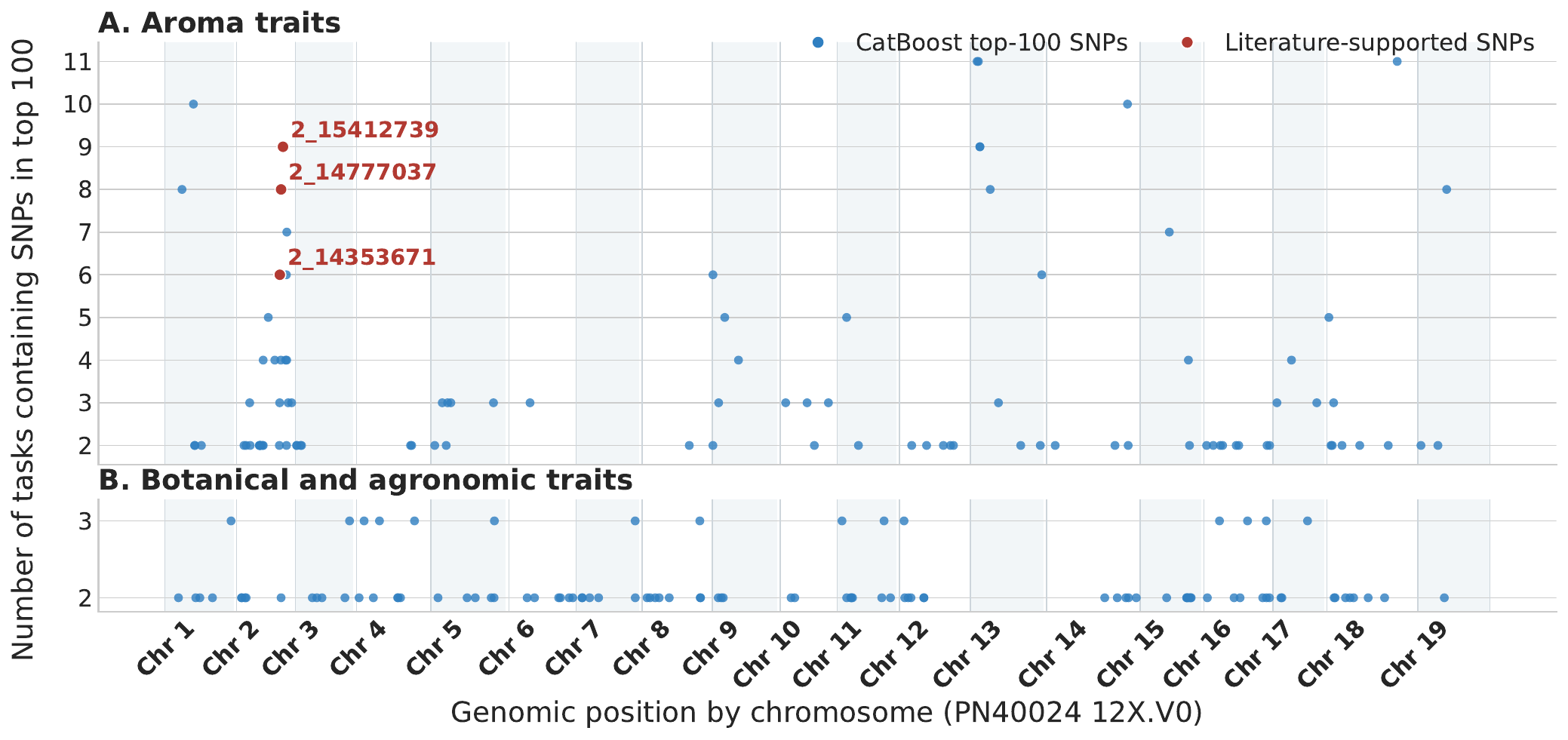}
	\caption{Recurrence of CB-prioritized SNPs across phenotype tasks. Each panel shows the 100 most frequent SNPs occurring in task-level top-100 lists. Red markers show literature-supported SNPs.}
    \label{fig:snp-recurrence}
\end{figure}

We performed a post-hoc analysis of the SNPs prioritized by the best-overall-performing model: CatBoost. For each phenotype task, SNPs were assigned a consensus rank using \texttt{PredictionValuesChange} feature importance. Figure~\ref{fig:snp-recurrence} shows the 100 most recurrent chromosomally-anchored SNPs within the aroma and botanical task families.
Three markers on chromosome 2 appeared among the top 100 markers for nine, eight, and six aroma tasks, respectively. These markers fall within the broader berry-colour region identified in previous linkage and QTL-mapping studies, which includes \textit{VvMYBA1} and \textit{VvMYBA2} \citep{grapecolor,berrytraits,garnachablanca}.
Botanical and agronomic tasks prioritized largely distinct SNP sets, with no marker recurring in more than three task-level top-100 lists. Once validated, these trait-specific candidates could potentially inform the breeding and selection of cultivars adapted to changing disease pressures or preferred growth characteristics.

\section{Conclusion}
To move toward a more quantitative understanding of how genetics influences grapevine phenotypes, we introduce a unified benchmark for sensory, botanical, and agronomic prediction. Climate change makes the choice of grapevine cultivars increasingly important. Botanical traits such as budburst timing, ripening period, acidity, sugar accumulation and disease resistance, shape how a cultivar performs under changing growing conditions. Understanding the genomic basis of these traits can potentially help identify plants that are better suited to changing temperatures, seasons, and disease pressures. Our benchmark provides a foundation to screen grapevine diversity more systematically in the future.

We demonstrate that classical tree-based models remain highly competitive for moderate-sample genomic prediction, acknowledging, though, that performance varies considerably across traits, with aroma and color generally proving more predictable than botanical and agronomic characteristics---adhering to grape-genomics literature. Within the matched Transformer comparison, our results suggest that learning a shared multi-task representation improves average predictive performance over separate single-task training. The primary limitation in our study, is the limited number of labeled cultivars for several phenotypes. Future work will expand phenotype coverage and annotation depth before testing generalization in an independent cultivar cohort. Moreover, our results should not be interpreted as identifying genomic markers from which plants with optimal adaptation to changing climatic conditions can be directly inferred. Rather, our aim is to demonstrate that genotype-to-phenotype prediction is feasible and can reveal predictive signals that domain experts might consider when investigating factors associated with plant resilience and environmental adaptation. Establishing whether such signals correspond to causal and transferable determinants of climate adaptation is left to future interdisciplinary work.

We hope that our benchmark will encourage oenologists, viticulturists, and climate experts to expand the dataset with additional cultivars and high-fidelity phenotype annotations. Ultimately, understanding genotype to phenotype interactions will be essential to select climate-resilient cultivars and adapt winemaking practices to preserve wine quality in a changing climate.

\bibliographystyle{plain}
\bibliography{biblio}

@article{geformer,
  title={GEFormer: A genotype-environment interaction-based genomic prediction method that integrates the gating multilayer perceptron and linear attention mechanisms},
  author={Yao, Zhou and Yao, Mengting and Wang, Chuang and Li, Ke and Guo, Junhao and Xiao, Yingjie and Yan, Jianbing and Liu, Jianxiao},
  journal={Molecular Plant},
  volume={18},
  number={3},
  pages={527--549},
  year={2025},
  publisher={Elsevier}
}

@article{cropformer,
  title={Cropformer: An interpretable deep learning framework for crop genomic prediction},
  author={Wang, Hao and Yan, Shen and Wang, Wenxi and Chen, Yongming and Hong, Jingpeng and He, Qiang and Diao, Xianmin and Lin, Yunan and Chen, Yanqing and Cao, Yongsheng and others},
  journal={Plant Communications},
  volume={6},
  number={3},
  year={2025},
  publisher={Elsevier}
}

@article{enformer,
  title={Effective gene expression prediction from sequence by integrating long-range interactions},
  author={Avsec, {\v{Z}}iga and Agarwal, Vikram and Visentin, Daniel and Ledsam, Joseph R and Grabska-Barwinska, Agnieszka and Taylor, Kyle R and Assael, Yannis and Jumper, John and Kohli, Pushmeet and Kelley, David R},
  journal={Nature methods},
  volume={18},
  number={10},
  pages={1196--1203},
  year={2021},
  publisher={Nature Publishing Group US New York}
}

@article{dnabert,
  author  = {Ji, Yanrong and Zhou, Zhihan and Liu, Han and Davuluri, Ramana V.},
  title   = {{DNABERT}: Pre-trained Bidirectional Encoder Representations from Transformers Model for {DNA}-Language in Genome},
  journal = {Bioinformatics},
  year    = {2021},
  volume  = {37},
  number  = {15},
  pages   = {2112--2120},
  doi     = {10.1093/bioinformatics/btab083},
  url     = {https://doi.org/10.1093/bioinformatics/btab083}
}

@book{wset,
  author    = {{Wine \& Spirit Education Trust}},
  title     = {Level 3 Award in Wines: Specification and Systematic Approach to Tasting Wine},
  year      = {2017},
  publisher = {WSET},
  address   = {London, UK},
  url       = {https://wsetglobal.com}
}

@inproceedings{attnnnn,
  title={Attention is All you Need},
  author={Ashish Vaswani and Noam Shazeer and Niki Parmar and Jakob Uszkoreit and Llion Jones and Aidan N. Gomez and Lukasz Kaiser and Illia Polosukhin},
  booktitle={Neural Information Processing Systems},
  year={2017},
  url={https://api.semanticscholar.org/CorpusID:13756489}
}

@article{climateaffectswine,
  title={Climate change impacts and adaptations of wine production},
  author={Van Leeuwen, Cornelis and Sgubin, Giovanni and Bois, Benjamin and Ollat, Nathalie and Swingedouw, Didier and Zito, S{\'e}bastien and Gambetta, Gregory A},
  journal={Nature Reviews Earth \& Environment},
  volume={5},
  number={4},
  pages={258--275},
  year={2024},
  publisher={Nature Publishing Group UK London}
}

@article{lacou,
  author = {Laucou, Valérie and Launay, Amandine and Bacilieri, Roberto
            and Lacombe, Thierry and Adam-Blondon, Anne-Françoise
            and Bérard, Aurélie and Chauveau, Aurélie
            and de Andrés, Maria Teresa and Hausmann, Ludger
            and Ibáñez, Javier and Le Paslier, Marie-Christine
            and Maghradze, David and Martinez-Zapater, José Miguel
            and Maul, Erika and Ponnaiah, Maharajah and Töpfer, Reinhard
            and Péros, Jean-Pierre and Boursiquot, Jean-Michel},
  title = {Extended diversity analysis of cultivated grapevine
           Vitis vinifera with 10K genome-wide SNPs},
  journal = {PLOS ONE},
  volume = {13},
  number = {2},
  pages = {e0192540},
  year = {2018},
  doi = {10.1371/journal.pone.0192540},
  url = {https://doi.org/10.1371/journal.pone.0192540}
}

@article{donofrio,
  title={Parentage atlas of Italian grapevine varieties as inferred from SNP genotyping},
  author={D’Onofrio, Claudio and Tumino, Giorgio and Gardiman, Massimo and Crespan, Manna and Bignami, Cristina and De Palma, Laura and Barbagallo, Maria Gabriella and Muganu, Massimo and Morcia, Caterina and Novello, Vittorino and others},
  journal={Frontiers in Plant Science},
  volume={11},
  pages={605934},
  year={2021},
  publisher={Frontiers Media SA}
}

@article{delorenzis,
  title={SNP genotyping elucidates the genetic diversity of Magna Graecia grapevine germplasm and its historical origin and dissemination},
  author={De Lorenzis, Gabriella and Mercati, Francesco and Bergamini, Carlo and Cardone, Maria Francesca and Lupini, Antonio and Mauceri, Antonio and Caputo, Angelo Raffaele and Abbate, Loredana and Barbagallo, Maria Gabriella and Antonacci, Donato and others},
  journal={BMC Plant Biology},
  volume={19},
  number={1},
  pages={7},
  year={2019},
  publisher={Springer}
}

@article{mlpinotnoirquality,
  title={Understanding Quality of Pinot Noir Wine: Can Modelling and Machine Learning Pave the Way?},
  author={Parul Tiwari and Piyush Bhardwaj and Sarawoot Somin and Wendy V. Parr and Roland Harrison and Don Kulasiri},
  journal={Foods},
  year={2022},
  volume={11},
  url={https://api.semanticscholar.org/CorpusID:252776426}
}

@Article{predictingwineunderchangingclimate,
AUTHOR = {Gačnik, Maja Borlinič and Škraba, Andrej and Pažek, Karmen and Rozman, Črtomir},
TITLE = {Predicting Wine Quality Under Changing Climate: An Integrated Approach Combining Machine Learning, Statistical Analysis, and Systems Thinking},
JOURNAL = {Beverages},
VOLUME = {11},
YEAR = {2025},
NUMBER = {4},
ARTICLE-NUMBER = {116},
URL = {https://www.mdpi.com/2306-5710/11/4/116},
ISSN = {2306-5710},
DOI = {10.3390/beverages11040116}
}

@misc{vivc,
  author = {Röckel, F. and Trapp, O. and Zyprian, E. and Hausmann, L. and Migliaro, D. and Vezzulli, S. and Töpfer, R. and Maul, E.},
  title = {Vitis International Variety Catalogue},
  howpublished = {\url{www.vivc.de}},
  note = {Accessed: August 2026}
}

@article{roshan2011ranking,
  title   = {Ranking causal variants and associated regions in
             genome-wide association studies by the support vector
             machine and random forest},
  author  = {Roshan, Usman and Chikkagoudar, Satish and Wei, Zhi
             and Wang, Kai and Hakonarson, Hakon},
  journal = {Nucleic Acids Research},
  volume  = {39},
  number  = {9},
  pages   = {e62},
  year    = {2011},
  doi     = {10.1093/nar/gkr064}
}

@article{mlgeneralwinequality,
  author  = {Khilari, Nitin and Hadawale, Pravin and Shaikh, Hasan and Kolase, Sachin},
  title   = {Analysis of Machine Learning Algorithm to Predict Wine Quality},
  journal = {International Journal of Scientific Research in Science, Engineering and Technology},
  year    = {2022},
  volume  = {9},
  number  = {2},
  pages   = {231--236},
  doi     = {10.32628/IJSRSET229235},
  url     = {https://doi.org/10.32628/IJSRSET229235}
}

@article{genophenotablegrapes,
    author = {Migicovsky, Zoë and Sawler, Jason and Gardner, Kyle M and Aradhya, Mallikarjuna K and Prins, Bernard H and Schwaninger, Heidi R and Bustamante, Carlos D and Buckler, Edward S and Zhong, Gan-Yuan and Brown, Patrick J and Myles, Sean},
    title = {Patterns of genomic and phenomic diversity in wine and table grapes},
    journal = {Horticulture Research},
    volume = {4},
    pages = {17035},
    year = {2017},
    month = {01},
    issn = {2052-7276},
    doi = {10.1038/hortres.2017.35},
    url = {https://doi.org/10.1038/hortres.2017.35},
    eprint = {https://academic.oup.com/hr/article-pdf/doi/10.1038/hortres.2017.35/41957796/41438_2017_article_bfhortres201735.pdf},
}

@misc{spawn99,
  author       = {spawn99},
  title        = {Wine Reviews Dataset},
  year         = {2025},
  publisher    = {Hugging Face},
  howpublished = {\url{https://huggingface.co/datasets/spawn99/wine-reviews}},
  note         = {Version 552d147}
}

@article{flutre2022prediction,
  author  = {Flutre, Timoth{\'e}e and Le Cunff, Lo{\"i}c
             and Fodor, Agota and others},
  title   = {A genome-wide association and prediction study
             in grapevine deciphers the genetic architecture
             of multiple traits and identifies genes under
             many new {QTLs}},
  journal = {G3: Genes, Genomes, Genetics},
  year    = {2022},
  volume  = {12},
  number  = {7},
  pages   = {jkac103},
  doi     = {10.1093/g3journal/jkac103}
}

@article{brault2021drought,
  author  = {Brault, Charlotte and Doligez, Agn{\`e}s
             and Le Cunff, Lo{\"i}c and Coupel-Ledru, Aude
             and Simonneau, Thierry and Chiquet, Julien
             and This, Patrice and Flutre, Timoth{\'e}e},
  title   = {Harnessing multivariate, penalized regression
             methods for genomic prediction and {QTL} detection
             of drought-related traits in grapevine},
  journal = {G3: Genes, Genomes, Genetics},
  year    = {2021},
  volume  = {11},
  number  = {9},
  pages   = {jkab248},
  doi     = {10.1093/g3journal/jkab248}
}

@article{bayeswine,
  author  = {Mercati, Francesco and De Lorenzis, Gabriella and Mauceri, Antonio and Zerbo, Marcello and Brancadoro, Lucio and D'Onofrio, Claudio and Morcia, Caterina and Barbagallo, Maria Gabriella and Bignami, Cristina and Gardiman, Massimo and de Palma, Laura and Ruffa, Paola and Novello, Vittorino and Crespan, Manna and Sunseri, Francesco},
  title   = {Integrated Bayesian Approaches Shed Light on the Dissemination Routes of the Eurasian Grapevine Germplasm},
  journal = {Frontiers in Plant Science},
  year    = {2021},
  volume  = {12},
  pages   = {692661},
  doi     = {10.3389/fpls.2021.692661},
  url     = {https://doi.org/10.3389/fpls.2021.692661}
}

@article{yangWineBehavior,
  author  = {Yang, Jiyun and Lee, Jeehyun},
  title   = {Current Research Related to Wine Sensory Perception Since 2010},
  journal = {Beverages},
  year    = {2020},
  volume  = {6},
  number  = {3},
  pages   = {47},
  doi     = {10.3390/beverages6030047},
  url     = {https://doi.org/10.3390/beverages6030047}
}

@article{garnachablanca,
  author  = {Royo, Carolina and Rodríguez-Lorenzo, Maite and Carbonell-Bejerano, Pablo and Mauri, Nuria and Cibriáin, Félix and Suberviola, J. and Sagüés, A. and Ibáñez, Javier and Martínez-Zapater, José M.},
  title   = {Characterization of Deletions Causing Berry-Color Variation in Garnacha and Tempranillo},
  journal = {Acta Horticulturae},
  year    = {2019},
  volume  = {1248},
  pages   = {463--470},
  doi     = {10.17660/ActaHortic.2019.1248.64},
  url     = {https://doi.org/10.17660/ActaHortic.2019.1248.64}
}

@article{grapecolor,
  title={New quantitative trait locus (QTLs) and candidate genes associated with the grape berry color trait identified based on a high-density genetic map},
  author={Sun, Lei and Li, Shenchang and Jiang, Jianfu and Tang, Xiaoping and Fan, Xiucai and Zhang, Ying and Liu, Jihong and Liu, Chonghuai},
  journal={BMC Plant Biology},
  volume={20},
  number={1},
  pages={302},
  year={2020},
  publisher={Springer}
}

@article{berrytraits,
  title={Genome-wide association study of berry-related traits in grape [Vitis vinifera L.] based on genotyping-by-sequencing markers},
  author={Guo, Da-Long and Zhao, Hui-Li and Li, Qiong and Zhang, Guo-Hai and Jiang, Jian-Fu and Liu, Chong-Huai and Yu, Yi-He},
  journal={Horticulture research},
  volume={6},
  year={2019},
  publisher={Oxford Academic}
}

@article{vanleeuwen2019adaptation,
  author  = {van Leeuwen, Cornelis and Destrac-Irvine, Agnès and Dubernet, Matthieu and Duchêne, Eric and Gowdy, Mark and Marguerit, Elisa and Pieri, Philippe and Parker, Amber and de Rességuier, Laure and Ollat, Nathalie},
  title   = {An Update on the Impact of Climate Change in Viticulture and Potential Adaptations},
  journal = {Agronomy},
  year    = {2019},
  volume  = {9},
  number  = {9},
  pages   = {514},
  doi     = {10.3390/agronomy9090514},
  url     = {https://doi.org/10.3390/agronomy9090514}
}

@article{wolkovich2018pinot,
  author  = {Wolkovich, Elizabeth M. and García de Cortázar-Atauri, Iñaki and Morales-Castilla, Ignacio and Nicholas, Kimberly A. and Lacombe, Thierry},
  title   = {From Pinot to Xinomavro in the World's Future Wine-Growing Regions},
  journal = {Nature Climate Change},
  year    = {2018},
  volume  = {8},
  pages   = {29--37},
  doi     = {10.1038/s41558-017-0016-6},
  url     = {https://doi.org/10.1038/s41558-017-0016-6}
}

\appendix
\section{Appendix}

\subsection{Dataset sources details for phenotypic traits}\label{phenotypes}
Phenotype information has been obtained---where available---for each cultivar, through various sources. An outline is reported in Table \ref{tab:web-data-sources}.

\begin{table*}[h]
\centering
\caption{Web resources}
\label{tab:web-data-sources}
\begin{tabularx}{\textwidth}{@{}p{0.26\textwidth}X@{}}
\toprule
\textbf{Data layer} & \textbf{Sources} \\
\midrule

Sensory vocabulary and profiles &
\href{https://www.wsetglobal.com/media/3119/wset_l3_wines_sat_en_jun-2016.pdf}{WSET Level 3 SAT}, and \href{https://wset.luksow.com/}{Level 2 reference table}. 
\href{https://courtofmastersommeliers.org/wp-content/uploads/2022/10/Grape-Markers.pdf}{Court of Master Sommeliers Grape Markers}\\

\addlinespace

Plant and botanical traits &
\href{https://www.plantgrape.fr/en/varieties/fruit-varieties/}{PlantGrape/IFV},
\href{https://grapevinevariety.com/}{Wine Australia Grape Variety Selector},
\href{https://www.vivairauscedo.com/en/product-index/}{Vivai Cooperativi Rauscedo},
\href{https://selections.entav-inra.fr/en}{ENTAV-INRA},
\href{https://vitisdb.it}{Italian Vitis Database}, and
\href{https://www.vinetowinecircle.com/en/castas/}{Vine to Wine Circle}. \\

\addlinespace

Cultivar identity, origin, targeted sensory enrichment &
\href{https://www.vivc.de/index.php?r=cultivarname\%2Findex}{VIVC},
\href{https://eng-vassal.montpellier.hub.inrae.fr/}{INRAE Vassal catalogue},
\href{https://www.wine-searcher.com/grape-varieties}{Wine-Searcher},
Italian Wine Central, Viniou, Wines of Portugal, Wines of Greece, Wein.plus, and VinoVoss. \\

\bottomrule
\end{tabularx}
\end{table*}

\textit{Aroma}. We constructed a controlled aroma vocabulary around the primary aroma families in the WSET \cite{wset} Level 3 Systematic Approach to Tasting Wine lexicon. WSET distinguishes primary aromas---associated with the grape and alcoholic fermentation---from secondary winemaking, and tertiary maturation aromas. The original expert descriptions contain a relatively fine-grained vocabulary of individual aroma descriptors. For consistency purposes, we retained 11 broad primary prediction targets: floral, citrus fruit, red fruit, green fruit, pungent spice, black fruit, stone fruit, herbaceous/herbal, mineral, dried/cooked fruit and tropical fruit. Descriptors corresponding to herbaceous and herbal characteristics were combined into a single task because of their close semantic overlap. Note that categories are not mutually exclusive: a single cultivar may exhibit multiple aroma families simultaneously, allowing the resulting formulation to reflect the structure of the underlying sensory data more closely than a single multiclass classification problem. The number of labeled cultivars differed between aroma tasks, ranging from $N=398$ for dried/cooked fruit to $N=472$ for floral aroma. 

\textit{Botanical and agronomic.}
To complement the sensory phenotype space, we included botanical and agronomic characteristics, distinct from sensory perception. The original heterogeneous botanical annotations were harmonized into standardized phenological and morphological traits by mapping to a structured common representation. This preprocessing was designed to retain the ordinal structure of traits wherever possible rather than reducing inherently ordered characteristics to unrelated categorical labels. The resulting set includes 14 ternary ordinal traits, each represented using three ordered levels corresponding to the original annotation scale (examples in Table \ref{tab:benchmark-tasks}).

\subsection{Technical details}

\subsubsection{Training} \label{training}
The tree-based models use fixed-capacity protocols without gradient-based early stopping: CatBoost is trained for 350 iterations at depth 5 ($\mathrm{lr}=0.05$); XGBoost grows 500 trees with max depth 6, $\mathrm{lr}=0.05$, and $0.8$ subsampling for both rows and columns; Random Forest trains 500 trees with $\sqrt{p}$ features per split. L2-regularized logistic regression performs grid search over $C \in \{0.01, 0.1, 1.0, 10.0, 100.0\}$ evaluated on the validation partition, selecting the value that maximizes AUROC (with balanced accuracy as tiebreaker). The Transformer configurations instead, are optimized with AdamW, with a learning rate of $\mathrm{lr}=5\!\times\!10^{-4}$ and weight decay of $1\!\times\!10^{-3}$, reduced via cosine annealing to $\mathrm{lr}_{\min}=2.5\!\times\!10^{-5}$ over $E_{\max}=45$ epochs with early stopping (patience=10, min. epochs=8). Mixed-precision (bfloat16) training was applied. For the multi-task Transformer, training examples are identity-task pairs, and a weighted sampler ensures each task receives equal expected sampling mass; checkpointing is driven by mean validation AUROC across estimable tasks. 

\paragraph{Genomic-prediction baselines.}\label{genomic-baselines}
GBLUP is fitted using \texttt{rrBLUP}, with a fixed intercept and variance components estimated by REML from the training data. Missing SNP dosages are replaced with training means. We exclude SNPs that are entirely missing or have only one allele in the
training data. 

For elastic net and LASSO, missing dosages are replaced with training means and SNPs are scaled by the standard deviations
of observed training calls. Constant and entirely missing SNPs are excluded. Class indicators are centered using their training
frequencies, which are included as unpenalized intercepts. Elastic net is tuned over L1 ratios of 0.1, 0.5 and 0.9, while LASSO uses an L1 ratio of 1. For each ratio, we test ten log-spaced penalties, starting at the threshold that sets all SNP coefficients to zero and decreasing to one-thousandth of that value. We select the model with the highest validation AUROC and use balanced accuracy to break ties.

\subsubsection{Training and evaluation specifics}\label{evaluation}
\textbf{Evaluation} For the Transformer models, the chi-square statistic is computed between each SNP locus and the task label; scores are weighted by the square root of per-marker call rate, and loci with fewer than 20 observations or insufficient class variation are filtered. The top 256 markers are retained independently per task and per fold, exclusively on the training partition. Tree-based baselines (CB, RF, XGB) and L2 logistic regression operate on the full set of 10,206 SNPs without pre-selection. This choice is tied to the models' intrinsic architecture. 
CB, RF and XGB train on the training partition only, with no inner loop. The validation partition is used by the Transformers for early stopping and by L2 logistic regression, elastic net and LASSO for hyperparameter selection. All models are evaluated independently on the held-out test partition, which sees no information during training, feature selection, or checkpoint choice. Outer evaluation metrics---macro one-vs-rest AUROC, balanced accuracy (BAC), and macro-F1---are computed on this test partition.
All models, except L2 logistic regression, GBLUP, elastic net and LASSO, are run with three base random seeds (42, 142, 242). Results are aggregated as means and standard deviations across folds and seeds. External-panel cultivars (D'Onofrio et al. and De Lorenzis et al. \cite{donofrio, delorenzis}) are restricted to the training partition; validation and test partitions contain cultivars from the Laucou et al. \cite{lacou} panel only, ensuring that held-out evaluation reflects generalization to unseen identities from the core reference panel (i.e.: the most complete).

\subsubsection{Transformer} \label{transformer}
The two Transformer models have been included in the benchmark to analyze whether the attention mechanism would have improved representation learning and modeling of intrinsic SNP data. Below we further describe architectural details inherent to the two relative configurations we present, and any additional technical detail that differs from the other ML models included in the study. 

\textbf{Token representation.} For each selected marker (j), the observed diploid genotype (e.g., AA, AG, GG) is mapped to an integer token using a categorical vocabulary ($\mathcal{V}$) fitted exclusively on the corresponding training fold. The vocabulary reserves indices ({0,1}) for padding ($\mathrm{PAD}$) and out-of-vocabulary states ($\mathrm{UNK}$). Genotype tokens are mapped through a learned embedding table ($\mathbf{E}{\mathrm{geno}} \in \mathbb{R}^{(V+2)\times d{\mathrm{model}}}$). To encode marker identity, we add a learned marker embedding indexed by the SNP's global catalog identifier (1–10,206) rather than its selection rank or local sequence position.
Genomic location is encoded through two additional learned embeddings: a chromosome embedding ($\mathbf{e}^{\mathrm{chr}}{\mathrm{chr}(j)}$) and a binned physical-position embedding ($\mathbf{e}^{\mathrm{pos}}{\lfloor \mathrm{pos}(j)/s \rfloor}$), with bin size (s=250,000). Thus, the initial representation of marker (j) is
\[ h_j = E_{\mathrm{geno}}(x_j) + E_{\mathrm{marker}}(j) + E_{\mathrm{chr}}(c_j) + E_{\mathrm{pos}}(p_j) + E_{\mathrm{place}}(l_j), \]where $x_j$ denotes the genotype token, $c_j$ the chromosome index, $p_j$ the 250-kbp position bin, and $l_j$ a placement embedding distinguishing canonical chromosomes, random scaffolds, unplaced loci, and plastid markers. All embeddings have dimension 96. A learned classification token is prepended to the resulting marker sequence.

\textbf{Transformer encoder.}
The sequence of k-selected SNPs is processed by three pre-normalized Transformer encoder layers, each containing four attention heads, a 384-dimensional feed-forward block (hidden dimension), GELU activation and dropout of 0.22 on attention and MLP outputs. The encoded marker representations are summarized using learned attention pooling. The pooled marker vector is concatenated with the final classification-token representation and passed through a two-layer classification head to produce task-specific logits.
    
\subsubsection{Ablations.} 

As mentioned in the main text, the two Transformer models operate on a set of selected SNP markers. Selection is performed independently per task and per fold to prevent label leakage.
We performed ablations on the number of selected features by comparing $k=256$ against $k=128$. Interestingly, the $k=128$ configuration does not strongly impact performance for the multi-task Transformer ($\Delta$AUROC $k_{256} \rightarrow k_{128} -0.001 $), while it improves the single-task scores ($\Delta$AUROC $k_{256}\rightarrow k_{128}: +0.009$ ). 

\begin{table}[ht]
    \centering
    \caption{Single-task (ST) and multi-task (MT) performance across different values of $k$ selected features.}
    \label{tab:st_mt_comparison}
    \begin{tabular}{c|cc|cc}
        \toprule
        \multirow{2}{*}{$k$} 
        & \multicolumn{2}{c|}{\textbf{Single-task}} 
        & \multicolumn{2}{c}{\textbf{Multi-task}} \\
        & BAC & AUROC & BAC & AUROC \\
        \midrule
        128  & 0.5285 & 0.6552 & \textbf{0.5327} & \textbf{0.6661} \\
        512  & 0.5106 & 0.6409 & \textbf{0.5275} & \textbf{0.6672} \\
        1024 & 0.4972 & 0.6226 & \textbf{0.5165} & \textbf{0.6651} \\
        2048 & 0.4762 & 0.6047 & \textbf{0.5006} & \textbf{0.6450} \\
        4000 & 0.4443 & 0.5697 & \textbf{0.4577} & \textbf{0.5841} \\
        \bottomrule
    \end{tabular}
\end{table}

\subsection{Task-performance comparison}
\begin{figure}[h]
    \centering
    \includegraphics[width=\linewidth]{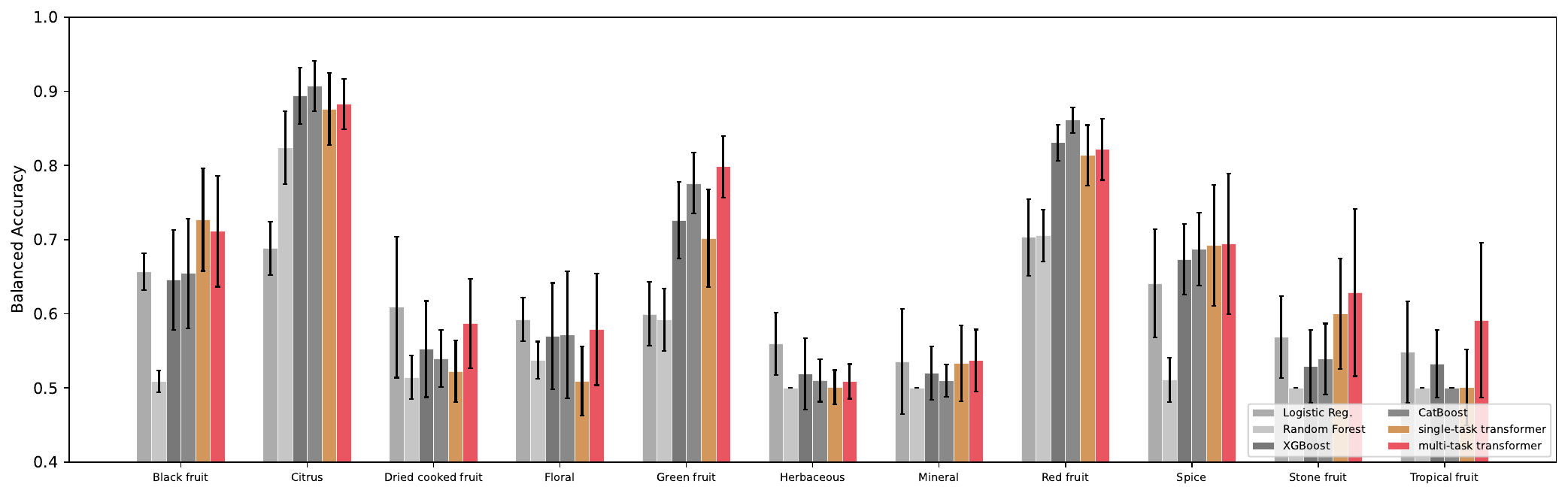}
    \caption{BAC on the 11 aroma-family tasks. L2-LR dominates on floral and herbaceous, while both transformer models outperform the others on black, stone and tropical fruit. Error bars represent per-fold standard deviation.}
    \label{fig:aromapred}
\end{figure}

\begin{figure}[h]
    \centering
    \includegraphics[width=\linewidth]{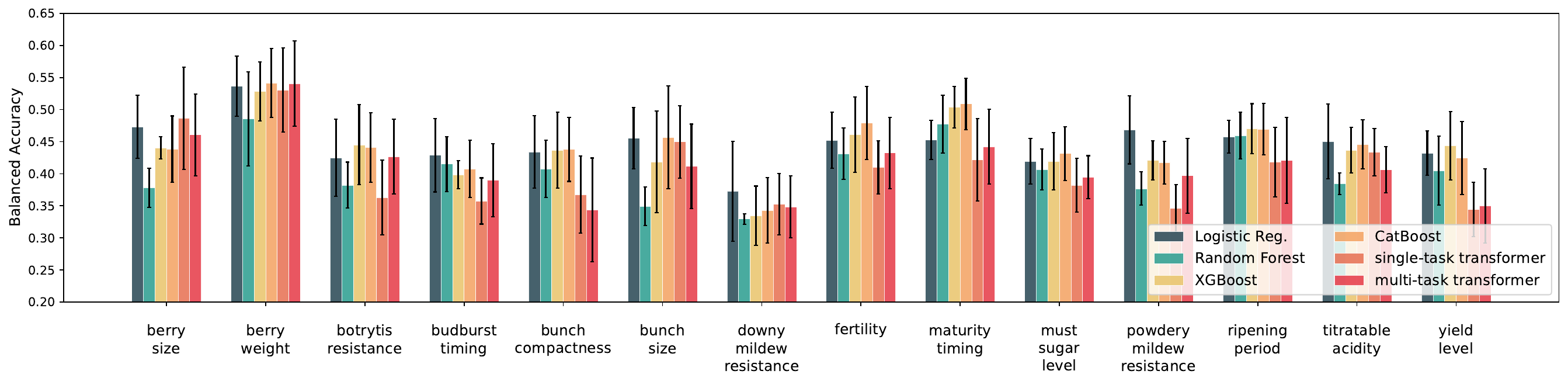}
    \caption{Balanced accuracy across the 14 botanical and agronomic prediction tasks. Per-fold standard deviations are shown as error bars.}
    \label{fig:botany_pred}
\end{figure}

\begin{figure}[h]
    \centering
    \includegraphics[width=0.6\linewidth]{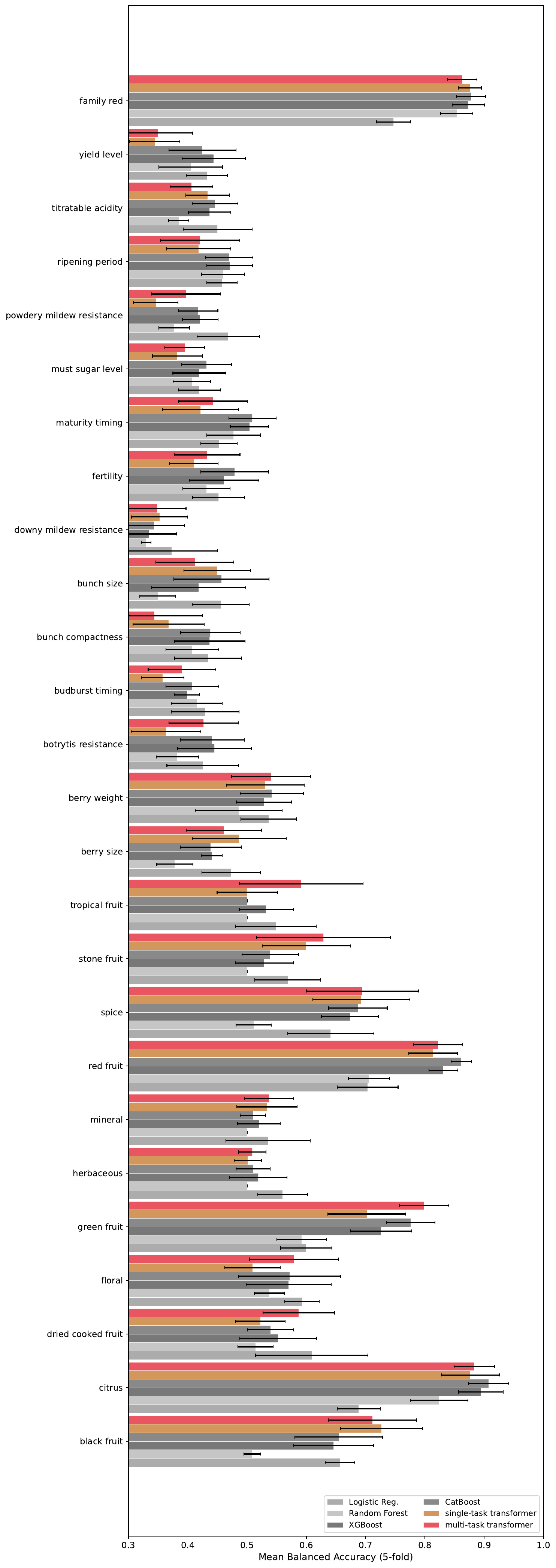}
    \caption{Task-level BAC for all 26 phenotype tasks and six model families. Tasks are grouped by phenotype class (color, aroma, then botanical) and sorted alphabetically within groups.}
    \label{fig:per_task_bac_models}
\end{figure}

\begin{figure}[h]
    \centering
    \includegraphics[width=\linewidth]{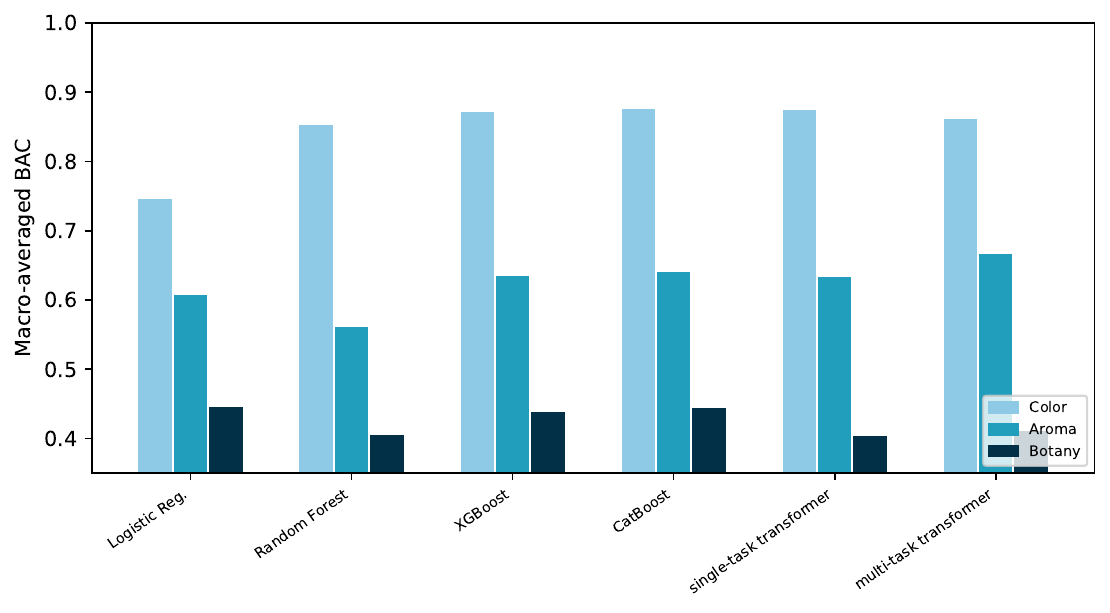}
    \caption{Macro-averaged BAC for each model, grouped by phenotype family. Performance drops consistently from color to aroma to botanical traits, reflecting increasing task difficulty.}
    \label{fig:macro_bac_models}
\end{figure}

\begin{figure}[h]
    \centering
    \includegraphics[width=\linewidth]{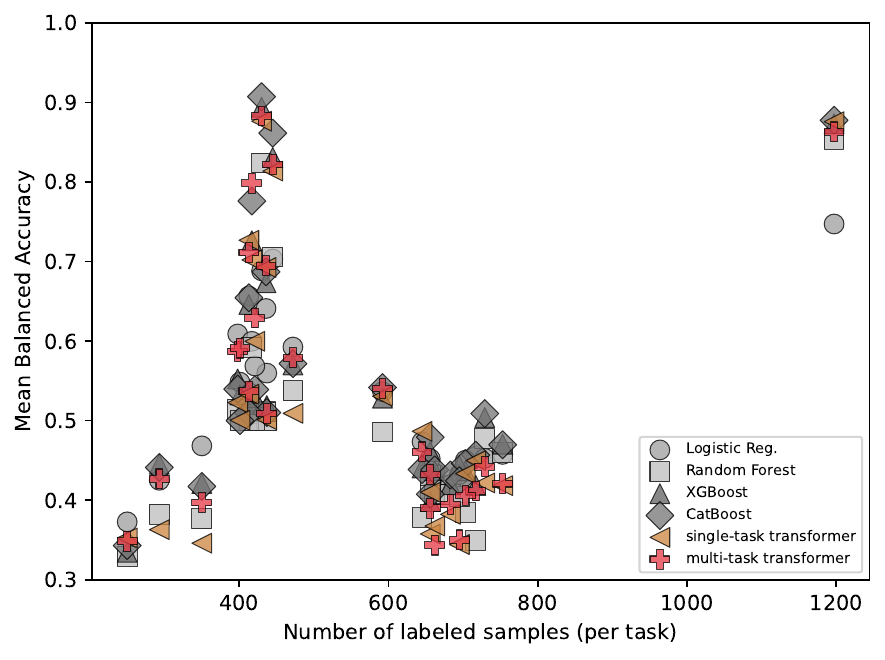}
    \caption{Mean BAC versus the number of labeled cultivars per task. Color-task points cluster at $N=1{,}197$ with high performance; aroma tasks ($N=398$--$472$) show variable predictability; botanical tasks ($N=590--750$) sit slightly above the aroma tasks.}
    \label{fig:bacvssamplesize}
\end{figure}

Figures \ref{fig:aromapred}, \ref{fig:botany_pred} report model performances across aroma and botany tasks---Figure \ref{fig:per_task_bac_models} reports each model's BAC performance per-task tasks, while Figure \ref{fig:macro_bac_models} refers macro-BAC. In order to analyze sample-size impact on predictions, Figure \ref{fig:bacvssamplesize} has been reported. 

\subsection{Complete list of task performances per model}

\begin{longtable}{lrrr}
\caption{CatBoost performance across phenotypic prediction tasks.}
\label{tab:catboost_results}\\
\toprule
Task & AUROC & BAC & Macro F1 \\
\midrule
\endfirsthead

\toprule
Task & AUROC & BAC & Macro F1 \\
\midrule
\endhead

\bottomrule
\endfoot

aroma black fruit & 0.871 & 0.655 & 0.665 \\
aroma citrus & 0.921 & 0.907 & 0.901 \\
aroma dried/cooked fruit & 0.651 & 0.540 & 0.548 \\
aroma floral & 0.619 & 0.572 & 0.570 \\
aroma green fruit & 0.837 & 0.776 & 0.766 \\
aroma herbaceous & 0.540 & 0.510 & 0.467 \\
aroma mineral & 0.631 & 0.510 & 0.480 \\
aroma red fruit & 0.890 & 0.861 & 0.843 \\
aroma spice & 0.820 & 0.687 & 0.694 \\
aroma stone fruit & 0.796 & 0.539 & 0.527 \\
aroma tropical fruit & 0.704 & 0.500 & 0.476 \\
botany berry size & 0.702 & 0.439 & 0.434 \\
botany berry weight & 0.740 & 0.542 & 0.542 \\
botany botrytis resistance & 0.655 & 0.441 & 0.435 \\
botany budburst timing & 0.623 & 0.408 & 0.399 \\
botany bunch compactness & 0.627 & 0.438 & 0.437 \\
botany bunch size & 0.654 & 0.457 & 0.465 \\
botany downy mildew resistance & 0.492 & 0.343 & 0.304 \\
botany fertility & 0.638 & 0.479 & 0.477 \\
botany maturity timing & 0.671 & 0.509 & 0.516 \\
botany must sugar level & 0.625 & 0.432 & 0.425 \\
botany powdery mildew resistance & 0.634 & 0.417 & 0.394 \\
botany ripening period & 0.639 & 0.470 & 0.477 \\
botany titratable acidity & 0.619 & 0.446 & 0.445 \\
botany yield level & 0.598 & 0.425 & 0.416 \\
color family red & 0.908 & 0.878 & 0.878 \\

\end{longtable}

\begin{longtable}{lrrr}
\caption{L2-regularized logistic regression performance across phenotypic prediction tasks.}
\label{tab:l2_logistic_results}\\
\toprule
Task & AUROC & BAC & Macro F1 \\
\midrule
\endfirsthead

\toprule
Task & AUROC & BAC & Macro F1 \\
\midrule
\endhead

\bottomrule
\endfoot

aroma black fruit & 0.725 & 0.657 & 0.659 \\
aroma citrus & 0.768 & 0.688 & 0.678 \\
aroma dried/cooked fruit & 0.613 & 0.609 & 0.584 \\
aroma floral & 0.636 & 0.593 & 0.590 \\
aroma green fruit & 0.690 & 0.600 & 0.595 \\
aroma herbaceous & 0.561 & 0.560 & 0.539 \\
aroma mineral & 0.605 & 0.535 & 0.528 \\
aroma red fruit & 0.747 & 0.703 & 0.701 \\
aroma spice & 0.679 & 0.641 & 0.634 \\
aroma stone fruit & 0.619 & 0.569 & 0.555 \\
aroma tropical fruit & 0.684 & 0.548 & 0.529 \\
botany berry size & 0.677 & 0.473 & 0.468 \\
botany berry weight & 0.707 & 0.536 & 0.517 \\
botany botrytis resistance & 0.646 & 0.425 & 0.425 \\
botany budburst timing & 0.610 & 0.429 & 0.424 \\
botany bunch compactness & 0.622 & 0.434 & 0.434 \\
botany bunch size & 0.625 & 0.455 & 0.457 \\
botany downy mildew resistance & 0.546 & 0.373 & 0.353 \\
botany fertility & 0.600 & 0.452 & 0.454 \\
botany maturity timing & 0.627 & 0.453 & 0.454 \\
botany must sugar level & 0.604 & 0.420 & 0.418 \\
botany powdery mildew resistance & 0.636 & 0.468 & 0.462 \\
botany ripening period & 0.615 & 0.458 & 0.456 \\
botany titratable acidity & 0.613 & 0.450 & 0.442 \\
botany yield level & 0.605 & 0.432 & 0.431 \\
color family red & 0.820 & 0.747 & 0.747 \\

\end{longtable}

\begin{longtable}{lrrr}
\caption{Multitask Transformer performance across phenotypic prediction tasks.}
\label{tab:multitask_transformer_results}\\
\toprule
Task & AUROC & BAC & Macro F1 \\
\midrule
\endfirsthead

\toprule
Task & AUROC & BAC & Macro F1 \\
\midrule
\endhead

\bottomrule
\endfoot

aroma black fruit & 0.846 & 0.711 & 0.690 \\
aroma citrus & 0.920 & 0.883 & 0.874 \\
aroma dried/cooked fruit & 0.614 & 0.587 & 0.598 \\
aroma floral & 0.602 & 0.579 & 0.555 \\
aroma green fruit & 0.822 & 0.799 & 0.781 \\
aroma herbaceous & 0.504 & 0.509 & 0.485 \\
aroma mineral & 0.666 & 0.537 & 0.522 \\
aroma red fruit & 0.890 & 0.822 & 0.816 \\
aroma spice & 0.784 & 0.695 & 0.686 \\
aroma stone fruit & 0.780 & 0.629 & 0.596 \\
aroma tropical fruit & 0.758 & 0.591 & 0.554 \\
botany berry size & 0.664 & 0.461 & 0.413 \\
botany berry weight & 0.702 & 0.540 & 0.515 \\
botany botrytis resistance & 0.622 & 0.427 & 0.410 \\
botany budburst timing & 0.561 & 0.390 & 0.335 \\
botany bunch compactness & 0.532 & 0.344 & 0.306 \\
botany bunch size & 0.624 & 0.412 & 0.386 \\
botany downy mildew resistance & 0.508 & 0.348 & 0.317 \\
botany fertility & 0.598 & 0.432 & 0.387 \\
botany maturity timing & 0.610 & 0.442 & 0.407 \\
botany must sugar level & 0.584 & 0.395 & 0.346 \\
botany powdery mildew resistance & 0.554 & 0.397 & 0.367 \\
botany ripening period & 0.595 & 0.421 & 0.385 \\
botany titratable acidity & 0.592 & 0.406 & 0.364 \\
botany yield level & 0.528 & 0.350 & 0.315 \\
color family red & 0.889 & 0.863 & 0.863 \\

\end{longtable}

\begin{longtable}{lrrr}
\caption{Random Forest performance across phenotypic prediction tasks.}
\label{tab:random_forest_results}\\
\toprule
Task & AUROC & BAC & Macro F1 \\
\midrule
\endfirsthead

\toprule
Task & AUROC & BAC & Macro F1 \\
\midrule
\endhead

\bottomrule
\endfoot

aroma black fruit & 0.760 & 0.509 & 0.459 \\
aroma citrus & 0.899 & 0.824 & 0.821 \\
aroma dried/cooked fruit & 0.669 & 0.514 & 0.503 \\
aroma floral & 0.643 & 0.538 & 0.459 \\
aroma green fruit & 0.785 & 0.592 & 0.579 \\
aroma herbaceous & 0.577 & 0.500 & 0.431 \\
aroma mineral & 0.611 & 0.500 & 0.456 \\
aroma red fruit & 0.839 & 0.706 & 0.710 \\
aroma spice & 0.715 & 0.511 & 0.446 \\
aroma stone fruit & 0.663 & 0.500 & 0.442 \\
aroma tropical fruit & 0.712 & 0.500 & 0.476 \\
botany berry size & 0.710 & 0.378 & 0.347 \\
botany berry weight & 0.748 & 0.486 & 0.485 \\
botany botrytis resistance & 0.653 & 0.382 & 0.339 \\
botany budburst timing & 0.629 & 0.415 & 0.391 \\
botany bunch compactness & 0.657 & 0.408 & 0.382 \\
botany bunch size & 0.665 & 0.349 & 0.288 \\
botany downy mildew resistance & 0.534 & 0.330 & 0.266 \\
botany fertility & 0.641 & 0.431 & 0.424 \\
botany maturity timing & 0.665 & 0.477 & 0.477 \\
botany must sugar level & 0.626 & 0.407 & 0.363 \\
botany powdery mildew resistance & 0.653 & 0.377 & 0.328 \\
botany ripening period & 0.661 & 0.460 & 0.459 \\
botany titratable acidity & 0.640 & 0.385 & 0.338 \\
botany yield level & 0.618 & 0.405 & 0.362 \\
color family red & 0.895 & 0.854 & 0.854 \\

\end{longtable}

\begin{longtable}{lrrr}
\caption{Transformer performance across phenotypic prediction tasks.}
\label{tab:transformer_results}\\
\toprule
Task & AUROC & BAC & Macro F1 \\
\midrule
\endfirsthead

\toprule
Task & AUROC & BAC & Macro F1 \\
\midrule
\endhead

\bottomrule
\endfoot

aroma black fruit & 0.843 & 0.727 & 0.685 \\
aroma citrus & 0.899 & 0.876 & 0.869 \\
aroma dried/cooked fruit & 0.665 & 0.523 & 0.508 \\
aroma floral & 0.525 & 0.509 & 0.370 \\
aroma green fruit & 0.804 & 0.702 & 0.694 \\
aroma herbaceous & 0.504 & 0.501 & 0.388 \\
aroma mineral & 0.553 & 0.533 & 0.502 \\
aroma red fruit & 0.868 & 0.814 & 0.804 \\
aroma spice & 0.778 & 0.693 & 0.678 \\
aroma stone fruit & 0.714 & 0.600 & 0.570 \\
aroma tropical fruit & 0.482 & 0.500 & 0.468 \\
botany berry size & 0.669 & 0.487 & 0.346 \\
botany berry weight & 0.703 & 0.531 & 0.464 \\
botany botrytis resistance & 0.582 & 0.363 & 0.277 \\
botany budburst timing & 0.549 & 0.358 & 0.220 \\
botany bunch compactness & 0.565 & 0.368 & 0.251 \\
botany bunch size & 0.640 & 0.450 & 0.349 \\
botany downy mildew resistance & 0.546 & 0.352 & 0.261 \\
botany fertility & 0.604 & 0.410 & 0.342 \\
botany maturity timing & 0.618 & 0.422 & 0.353 \\
botany must sugar level & 0.566 & 0.382 & 0.290 \\
botany powdery mildew resistance & 0.534 & 0.346 & 0.217 \\
botany ripening period & 0.591 & 0.418 & 0.347 \\
botany titratable acidity & 0.599 & 0.433 & 0.360 \\
botany yield level & 0.516 & 0.344 & 0.228 \\
color family red & 0.895 & 0.876 & 0.876 \\

\end{longtable}

\begin{longtable}{lrrr}
\caption{XGBoost performance across phenotypic prediction tasks.}
\label{tab:xgboost_results}\\
\toprule
Task & AUROC & BAC & Macro F1 \\
\midrule
\endfirsthead

\toprule
Task & AUROC & BAC & Macro F1 \\
\midrule
\endhead

\bottomrule
\endfoot

aroma black fruit & 0.850 & 0.646 & 0.656 \\
aroma citrus & 0.910 & 0.894 & 0.889 \\
aroma dried/cooked fruit & 0.626 & 0.552 & 0.563 \\
aroma floral & 0.618 & 0.570 & 0.569 \\
aroma green fruit & 0.828 & 0.726 & 0.725 \\
aroma herbaceous & 0.547 & 0.519 & 0.499 \\
aroma mineral & 0.641 & 0.520 & 0.499 \\
aroma red fruit & 0.887 & 0.831 & 0.823 \\
aroma spice & 0.809 & 0.673 & 0.681 \\
aroma stone fruit & 0.767 & 0.529 & 0.520 \\
aroma tropical fruit & 0.721 & 0.532 & 0.534 \\
botany berry size & 0.694 & 0.440 & 0.440 \\
botany berry weight & 0.739 & 0.528 & 0.536 \\
botany botrytis resistance & 0.652 & 0.445 & 0.440 \\
botany budburst timing & 0.614 & 0.398 & 0.394 \\
botany bunch compactness & 0.619 & 0.437 & 0.438 \\
botany bunch size & 0.644 & 0.419 & 0.421 \\
botany downy mildew resistance & 0.488 & 0.335 & 0.300 \\
botany fertility & 0.621 & 0.461 & 0.459 \\
botany maturity timing & 0.665 & 0.504 & 0.511 \\
botany must sugar level & 0.620 & 0.420 & 0.412 \\
botany powdery mildew resistance & 0.621 & 0.421 & 0.403 \\
botany ripening period & 0.647 & 0.470 & 0.475 \\
botany titratable acidity & 0.613 & 0.437 & 0.438 \\
botany yield level & 0.615 & 0.444 & 0.437 \\
color family red & 0.898 & 0.873 & 0.873 \\

\end{longtable}

\begin{longtable}{lrrr}
\caption{Elastic net performance across phenotypic prediction tasks.}
\label{tab:elastic_net_results}\\
\toprule
Task & AUROC & BAC & Macro F1 \\
\midrule
\endfirsthead

\toprule
Task & AUROC & BAC & Macro F1 \\
\midrule
\endhead

\bottomrule
\endfoot

aroma black fruit & 0.819 & 0.702 & 0.711 \\
aroma citrus & 0.918 & 0.892 & 0.881 \\
aroma dried/cooked fruit & 0.662 & 0.531 & 0.532 \\
aroma floral & 0.609 & 0.578 & 0.556 \\
aroma green fruit & 0.832 & 0.798 & 0.778 \\
aroma herbaceous & 0.568 & 0.550 & 0.532 \\
aroma mineral & 0.574 & 0.507 & 0.478 \\
aroma red fruit & 0.884 & 0.861 & 0.839 \\
aroma spice & 0.825 & 0.694 & 0.688 \\
aroma stone fruit & 0.785 & 0.495 & 0.440 \\
aroma tropical fruit & 0.677 & 0.565 & 0.545 \\
botany berry size & 0.666 & 0.367 & 0.325 \\
botany berry weight & 0.721 & 0.406 & 0.344 \\
botany botrytis resistance & 0.643 & 0.433 & 0.420 \\
botany budburst timing & 0.592 & 0.416 & 0.408 \\
botany bunch compactness & 0.591 & 0.399 & 0.384 \\
botany bunch size & 0.644 & 0.338 & 0.261 \\
botany downy mildew resistance & 0.521 & 0.322 & 0.287 \\
botany fertility & 0.602 & 0.377 & 0.342 \\
botany maturity timing & 0.655 & 0.440 & 0.422 \\
botany must sugar level & 0.597 & 0.379 & 0.367 \\
botany powdery mildew resistance & 0.610 & 0.397 & 0.372 \\
botany ripening period & 0.635 & 0.445 & 0.428 \\
botany titratable acidity & 0.603 & 0.383 & 0.339 \\
botany yield level & 0.580 & 0.415 & 0.404 \\
color family red & 0.895 & 0.880 & 0.880 \\

\end{longtable}

\begin{longtable}{lrrr}
\caption{GBLUP performance across phenotypic prediction tasks.}
\label{tab:gblup_results}\\
\toprule
Task & AUROC & BAC & Macro F1 \\
\midrule
\endfirsthead

\toprule
Task & AUROC & BAC & Macro F1 \\
\midrule
\endhead

\bottomrule
\endfoot

aroma black fruit & 0.722 & 0.642 & 0.660 \\
aroma citrus & 0.771 & 0.693 & 0.680 \\
aroma dried/cooked fruit & 0.662 & 0.514 & 0.503 \\
aroma floral & 0.617 & 0.559 & 0.509 \\
aroma green fruit & 0.694 & 0.624 & 0.621 \\
aroma herbaceous & 0.569 & 0.521 & 0.496 \\
aroma mineral & 0.598 & 0.507 & 0.470 \\
aroma red fruit & 0.738 & 0.705 & 0.702 \\
aroma spice & 0.683 & 0.636 & 0.634 \\
aroma stone fruit & 0.638 & 0.502 & 0.471 \\
aroma tropical fruit & 0.666 & 0.500 & 0.476 \\
botany berry size & 0.707 & 0.389 & 0.361 \\
botany berry weight & 0.743 & 0.515 & 0.526 \\
botany botrytis resistance & 0.667 & 0.442 & 0.422 \\
botany budburst timing & 0.629 & 0.370 & 0.322 \\
botany bunch compactness & 0.652 & 0.349 & 0.247 \\
botany bunch size & 0.675 & 0.353 & 0.289 \\
botany downy mildew resistance & 0.561 & 0.350 & 0.298 \\
botany fertility & 0.638 & 0.461 & 0.454 \\
botany maturity timing & 0.676 & 0.472 & 0.476 \\
botany must sugar level & 0.610 & 0.380 & 0.313 \\
botany powdery mildew resistance & 0.649 & 0.394 & 0.353 \\
botany ripening period & 0.665 & 0.470 & 0.468 \\
botany titratable acidity & 0.639 & 0.386 & 0.342 \\
botany yield level & 0.612 & 0.399 & 0.347 \\
color family red & 0.825 & 0.759 & 0.758 \\

\end{longtable}

\begin{longtable}{lrrr}
\caption{LASSO performance across phenotypic prediction tasks.}
\label{tab:lasso_results}\\
\toprule
Task & AUROC & BAC & Macro F1 \\
\midrule
\endfirsthead

\toprule
Task & AUROC & BAC & Macro F1 \\
\midrule
\endhead

\bottomrule
\endfoot

aroma black fruit & 0.819 & 0.708 & 0.715 \\
aroma citrus & 0.919 & 0.905 & 0.898 \\
aroma dried/cooked fruit & 0.660 & 0.546 & 0.561 \\
aroma floral & 0.614 & 0.577 & 0.563 \\
aroma green fruit & 0.841 & 0.791 & 0.770 \\
aroma herbaceous & 0.556 & 0.520 & 0.508 \\
aroma mineral & 0.549 & 0.512 & 0.486 \\
aroma red fruit & 0.868 & 0.838 & 0.817 \\
aroma spice & 0.827 & 0.704 & 0.698 \\
aroma stone fruit & 0.789 & 0.497 & 0.450 \\
aroma tropical fruit & 0.664 & 0.515 & 0.496 \\
botany berry size & 0.659 & 0.388 & 0.355 \\
botany berry weight & 0.712 & 0.420 & 0.369 \\
botany botrytis resistance & 0.640 & 0.434 & 0.432 \\
botany budburst timing & 0.589 & 0.425 & 0.411 \\
botany bunch compactness & 0.600 & 0.409 & 0.406 \\
botany bunch size & 0.635 & 0.340 & 0.267 \\
botany downy mildew resistance & 0.491 & 0.337 & 0.284 \\
botany fertility & 0.596 & 0.380 & 0.344 \\
botany maturity timing & 0.648 & 0.436 & 0.425 \\
botany must sugar level & 0.591 & 0.379 & 0.369 \\
botany powdery mildew resistance & 0.613 & 0.390 & 0.368 \\
botany ripening period & 0.630 & 0.437 & 0.426 \\
botany titratable acidity & 0.607 & 0.401 & 0.384 \\
botany yield level & 0.577 & 0.403 & 0.389 \\
color family red & 0.895 & 0.877 & 0.877 \\

\end{longtable}

\clearpage
\end{document}